**Magnetic and transport anomalies in R$_2$RhSi$_3$ (R= Gd, Tb, and Dy) resembling those of an exotic magnetic skyrmion Gd$_2$PdSi$_3$**

Ram Kumar,[1] Kartik K Iyer,[1] P.L. Paulose[1] and E.V. Sampathkumaran[1,2]

[1]*Tata Institute of Fundamental Research, Homi Bhabha Road, Colaba, Mumbai 400005, India*
[2]*UGC-DAE CSR, Mumbai Centre, BARC Campus, Trombay, Mumbai 400085, India*

We have carried out magnetization, heat-capacity, electrical and magnetoresistance measurements (2-300 K) for the polycrystalline form of intermetallic compounds, R$_2$RhSi$_3$ (R= Gd, Tb and Dy), forming in a AlB$_2$-derived hexagonal structure with a triangular R network. This work was primarily motivated by a revival of interest on Gd$_2$PdSi$_3$ after about two decades in the field of Topological Hall Effect due to magnetic skyrmions. We report here that these compounds are characterized by double antiferromagnetic transitions ($T_N$= 13.5 and 12 K for Gd, 13.5 and 6.5 K for Tb; 6.5 and 2.5 K for Dy), but antiferromagnetism seems to be quite complex. The most notable observations common to all these compounds are: (i) There are many features in the data mimicking those seen for Gd$_2$PdSi$_3$, including the two field-induced changes in isothermal magnetization as though there are two metamagnetic transitions well below $T_N$. In view of such a resemblance of the properties, we speculate that these Rh-based materials offer a good playground to study topological Hall effect in a centrosymmetric structure, with its origin lying in triangular lattice of magnetic *R* ions; (ii) There is an increasing contribution of electronic scattering with decreasing temperature towards $T_N$ in all cases, similar to Gd$_2$PdSi$_3$, thereby serving as examples for a theoretical prediction for a classical spin-liquid phase in metallic magnetic systems due to geometrical frustration.

# I. INTRODUCTION

The area of research of 'magnetic skyrmions' characterized by vortex-like nanometric spin textures is one of the most exciting topics in condensed matter physics in the current literature due to potential spintronic applications as well as next generation information storage devices. A hallmark of such systems is that there is an additional contribution [1] to Hall resistivity in the magnetic field and temperature range in which such a magnetic texture crystallizes. The area of 'topological' Hall Effect (THE) in the context of magnetic skyrmions got triggered by the observation of a such phenomenon in an intermediate field range in MnSi and $Fe_{0.5}Co_{0.5}Si$ about a decade ago [2, 3]. We like to mention that an unexplainable large Hall resistivity was reported as early as 1999 on a compound $Gd_2PdSi_3$ [Refs. 4-6], particularly in an intermediate field range (following metamagnetic transitions) [7]. This work served as the key result to search for 'magnetic skyrmions' by resonant x-ray scattering in this compound successfully by Kurumaji et al [8]. An intriguing aspect of this observation of magnetic skyrmions is that this compound forms in a $AlB_2$-derived hexagonal (Fig. 1) (Refs. 9-11) centrosymmetric structure, emphasizing an unusual role of geometrically frustrated magnetism on the formation of magnetic skyrmions, instead of more commonly known role of Dzyaloshinskii-Moriya Interaction (DMI).

We had earlier provided sufficient evidence for the existence of both antiferromagnetic and ferromagnetic correlations along with many other magnetic, thermal and transport anomalies on $Gd_2PdSi_3$ [5-7, 12, 13]. In fact, the observation of [5] Kondo-like electrical resistivity ($\rho$) minimum before the onset of long range magnetic order – unexpected for heavy rare-earths like Gd – led to a novel theoretical advancement recently in magnetism, while we raised a question whether it is a signature of the formation of independent magnetic skyrmions in the paramagnetic state [4]; that is, Wang et al [14] have advanced a theory that such a minimum can also arise – not necessarily from the Kondo effect, but - from an interplay between Ruderman-Kittel-Kasuya-Yosida (RKKY) interaction and magnetic frustration due to geometrical arrangement of magnetic ions, even in the dilute limit. We therefore proposed [4] that it is of great interest to investigate other isostructural rare-earth (R) families to identify compounds with similar magnetic anomalies for a systematic understanding of such novel concepts, particularly to search for topological Hall anomalies due to magnetic skyrmions, and to throw open more materials with geometrical frustration in a centrosymmetric structure. We made exhaustive studies on entire Pd and other transition-metal (TM) based series, reporting many exotic features [see, for instance, Refs. 15-17, and articles cited therein]. Surprisingly, very little work has been done in the past literature on the Rh-based series, though some of the Ce compounds (i.e., $Ce_2RhSi_3$) and their solid solutions [18-24] attracted an attention due to an interplay between competition between magnetic ordering and the Kondo effect. For the benefit of the reader, we state that 46 structure types of binary and ternary intermetallic compounds have been derived from the well-known $AlB_2$ structure [25], and therefore it would be rewarding to probe this rich family in depth.

With this primary motivation, we have undertaken magnetic investigations on the three compounds in the Rh family, viz., $R_2RhSi_3$ (R= Gd, Tb and Dy), as a continuation of our ongoing efforts on such ternary families. Though these compounds form in the $AlB_2$-derived structure, it is not easy to resolve the space group, whether it is $P\bar{6}2c$ or $P6_3/mmc$ [Refs. 9, 11, 26] due to the fact that the atomic positions are very close for these space groups. Therefore, the deviation from the centrosymmetry for the former acentric space group is negligible [11]. Though these were studied by magnetization (*M*) in 1984 by the group which pioneered [9] the discovery of this Rh-family, it is difficult to infer subtle magnetic anomalies under discussion from the reported density of the data points. In that sense, the present studies are very exhaustive, particularly with the addition of data from other bulk methods. We report here new features, bringing out many commonalities in the magnetic and transport properties of these compounds. The points to be stressed are that the properties, in particular the observation of two metamagnetic transitions with ferromagnetic correlations competing with antiferromagnetism (AFM) and excess $\rho$ (as inferred from the suppression by external magnetic fields, $H$) are found to be qualitatively the same as those observed for $Gd_2PdSi_3$. We therefore wonder whether these compounds are potential candidates to search for other phenomenon, viz., topological Hall effect / magnetic skyrmions due to geometrical frustration characterizing this Pd-compound, in particular in an intermediate magnetic field range. Though this speculation is not straightforward, the field-induced metamagnetic transitions and THE are closely linked in other commonly known DMI-based magnetic skyrmions [1-3].

## II. EXPERIMENTAL DETAILS

The samples in polycrystalline form were prepared by arc melting stoichiometric amounts of high purity constituent elements (R >99.9%; Rh > 99.99%; Si >99.999%) by arc melting in an atmosphere of argon. The ingots were subsequently annealed at 1073 K for one week and characterized by x-ray diffraction ($K_\alpha$). Rietveld refined diffraction patterns along with fitted parameters are shown in Fig. 2; from a careful comparison of the goodness of the fits for the two space groups, we infer that these compounds may belong to $P6_3/mmc$ rather than $P\bar{6}2c$. The fits clearly revealed that there is a doubling of unit-cell parameters along $a$- and $c$- axis with respect to $AlB_2$ structure. Scanning electron microscopic images revealed homogeneity of the ingots and energy dispersive x-ray analysis confirmed that the composition almost corresponds to 2:1:3 (R:Rh:Si). Magnetization measurements as a function of temperature ($T$) and $H$ were carried out with the help of a commercial (Quantum Design) superconducting quantum interference device magnetometer. Heat-capacity ($C$), dc electrical resistance and magnetoresistance (MR) measurements by four-probe method were performed with the Physical Properties Measurements System (Quantun Design) down to 1.8 K. Unless stated, all the measurements were performed for the zero-field-cooled (ZFC) condition (to 2 K) of the specimens and the data were collected while warming.

## III. RESULTS AND DISCUSSION
### A. Magnetic susceptibility

Results of magnetic susceptibility ($\chi$) are shown in Fig. 3 for all the three samples in the form of $\chi$ versus $T$ (below 40 K in $H$= 100 Oe) and inverse $\chi$ versus $T$ (2-300 K in $H$= 5 kOe). We have also measured in a low-field of 100 Oe for field-cooled (FC) conditions; we did not observe any difference between ZFC and FC curves, attributable to spin-glass freezing (and hence FC curves are not shown). There is a well-defined peak in $\chi(T)$ at ($T_N$=) 13.5, 13.5 and 6.5 K for Gd, Tb and Dy cases respectively, indicating the onset of long-range AFM order. With the de Gennes factor being 0.667 and 0.45 for Tb and Dy with respect to Gd, there is a clear breakdown of de Gennes scaling for Tb, suggesting the role of 4f-anisotropy on magnetic ordering for Tb magnetism [27]. A careful look at the curves reveal, as inferred also from the derivative curves (for Gd) - not shown for the sake of clarity - and shoulders (for Tb and Dy), that there is an additional magnetic anomaly at 12, 6.5 and 2.5 K respectively. [In the case of Tb, a weak anomaly appears close to 4 K, as indicated by a vertical arrow; it is not clear whether it is intrinsic to the compound as heat-capacity data discussed below does not show a prominent feature. Possibly, a trace of some other impurity not detected by XRD may also be responsible for this feature]. It may be recalled [9, 11, 28] that the superstructure formation due to TM-Si crystallographic order, as shown in Fig. 1, results in a honeycomb network, made up of TM-Si hexagons and Si-Si hexagons with intercalation of R ions between hexagons. As a result, two sites for R are created, with the one at 2(b) site being immediately surrounded by 12 Si atoms and the other at 6(h) site by 4Rh atoms and 8 Si atoms. It is therefore not clear whether the two prominent magnetic features arise from these two types of R ions or whether there is a crossover from one type of AFM to another at the second feature. This ambiguity is however not relevant for the main conclusions of this article. Such double magnetic transitions were inferred [5, 6] also from Mössbauer spectroscopy, for $Gd_2PdSi_3$. A notable point is that the triangular network of R ions sandwitched between Si-Si (and Si-Rh) layers, though frustrated geometrically due to antiferromagnetic interaction, does not result in spin-glass freezing - a behavior similar to $Gd_2PdSi_3$. This is distinctly different from the spin-glass features observed for isostructural U compounds, in particular $U_2RhSi_3$ [Ref. 29]. With respect to inverse $\chi(T)$, there is a deviation below about 50 K from the high temperature linear behavior (well above $T_N$ notable clearly for Gd and Tb in Fig.3), which could be due to the classical spin-liquid phase (Ref. 14) in the paramagnetic state before long range magnetic order. Viewing this feature together with the two magnetic transitions mentioned above suggest that multiple magnetic phases compete in the event of geometrical frustration – an inference made on other geometrically frustrated families [30-33]. Further corroborative features are observed in the transport data (*vide infra*). The fact that magnetic frustration exists can be inferred from the values of paramagnetic Curie temperature ($\theta_p$) derived from the high temperature Curie-Weiss region (>100 K), see Fig. 3 (right); that is, these magnitudes are found to be significantly larger (~ -27 K, ~ -55 K and ~ - 12 K respectively) compared to respective $T_N$, as expected for magnetic frustration. The fact that the sign of $\theta_p$ is negative implies dominant antiferromagnetic correlations. Finally, the effective moment obtained from the Curie-Weiss region are found to be 8.1 and 10.9 $\mu_{eff}$/rare-earth for Gd and Dy in close agreement with free ion values. However, in the case of Tb, the value (10.4$\mu_B$/Tb) is found to be higher than the theoretical value for $Tb^{3+}$ (9.72 $\mu_B$). The reason for this enhanced value for Tb case is not clear to us at present; possibly, anisotropic Tb 4f-hybridization [27, 34] polarizes Rh 4d band.

### B. Heat-capacity

We report the results of zero-field and in-field heat-capacity measurements in Fig. 4a. It is clear that there is a well-defined λ-anomaly (and a peak) at the respective $T_N$ (inferred from the $\chi$ data above). Another feature is noted at a lower temperature – exactly at the same temperature where the second magnetic transition is inferred from the $\chi(T)$ data. A field of 10 kOe does not cause a shift of the peak temperature. However, with the application of higher magnetic fields, the peaks are gradually suppressed towards lower temperatures, which endorses that both the transitions in all cases are of antiferromagnetic types. An observation of interest is that the peak values get depressed with increasing *H*, particularly beyond 10 kOe, which is a signature of subtle changes in the modulation of AFM structure. In fact, for Gd systems, there are predictions what the peak value should be for a perfect equal-moment structure [35], according to which it should be 27 J/Gd mol. In contrast to this expectation, in $Gd_2RhSi_3$, the observed peak value (about 15 J/Gd mol) is far less establishing a complex modulated magnetic structures even in zero field. At very low temperatures, the $T^3$-form (in zero- field curves) expected for antiferromagnets is replaced by essentially linear form, suggesting complexity of AFM ground state. We have also derived magnetic contribution ($C_m$) to $C$ employing the $C$ values of La analogue using the procedure in Ref. 35 and it is found that there is a tail extending over a wide $T$-range above $T_N$ in all cases (see Fig. 4b row), establishing the existence of magnetic correlations before the onset of long range magnetic order. This even manifests itself as a broad peak about 15 K for the Dy case, and it is of interest to focus future studies on this aspect. Isothermal entropy change ($-\Delta S$), a measure of magnetocaloric effect was also derived (Fig. 4c rows). The $-\Delta S$ curves are found to lie in the positive quadrant initially (that is, at $T_N$) even for a field as low as 10 kOe, as a signature of a competition with ferromagnetic component with the presence of $H$, but the curves switch back to negative quadrant at lower temperatures for Gd and Tb cases, characteristic of dominating AF component. With increasing $H$, the curves stay back in the positive quadrant, as though field-induced ferromagnetic alignment dominates at higher fields. All these features are qualitatively the same as those known for $Gd_2PdSi_3$ [5, 13]. It is also obvious from the figures that the magnitude of $-\Delta S$ at the peak is rather large in general. For instance, for the Gd case, it is about 8 J/kg K, comparable to that for $Gd_2PdSi_3$ (for $H$= 0 to 50 kOe).

### C. Isothermal magnetization and magnetoresistance

In order to address the field-induced changes, we have obtained isothermal magnetization and magnetoresistance curves at selected temperatures below $T_N$. The curves are found to be essentially reversible with the variation of $H$ (that is, negligible hysteresis). The most relevant observation to the aim of this paper is that, at the lowest temperature measured (2 K), there is a significant decrease in the slopes of $M(H)$ curves near two fields, near 10 and 60 kOe for Gd, 12 and 38 kOe for Tb, and 6 and 60 kOe for Dy (Fig. 5). These are more clearly inferred from the derivative (d$M$/d$H$ versus $H$) curves (Fig. 5, inset). These metamagnetic-like transitions gradually get smeared with increasing $T$, and in the $T$-range between two AFM transitions, these are further smoothened (and hence not shown for the sake of clarity of figures). In consistence with these observations, MR [defined as $\{\rho(H)-\rho(0)\}/\rho(0)$] curves also reveal notable changes in slopes well below respective $T_N$, as shown in Fig. 6. The curves are symmetric with respect to origin. The magnitudes of MR are large, for instance, about -6% at 2 K in a field of 100 kOe. The sign of MR is negative even at low fields before field-induced transitions, rather than being positive as expected for an ideal antiferromagnet, and this implies the existence of magnetic super-zone gaps and/or the complexity of magnetic structure. At higher fields, negative sign is explainable due to gradual dominance of ferromagnetic interaction. All these features are comparable to those observed [5] for $Gd_2PdSi_3$, in particular for $H$//[0001] (Ref. 7).

### D. Temperature dependence of zero-field and in-field electrical resistivity

We show the results of zero-field and in-field electrical resistivity measurements as a function of temperature in Fig. 7. *We first discuss the data at $T_N$ and below.* It is obvious that, in all cases, there are anomalies at the onset of long range magnetic ordering in the zero-field curve, though these are manifested differently in these three compounds, following a gradual decrease of $\rho$ with $T$ in the paramagnetic state. In the case of Gd, there is a drop in $\rho$ at $T_N$ due to the loss of spin-disorder contribution. However, there is an initial upturn in the case of Tb and Dy attributable to magnetic Brillioun-zone boundary gaps. Such a behavior of $\rho(T)$ – that is, the suppression of the drop – was observed for $Gd_2PdSi_3$ as well. Clearly, these offer further support to the complexity of antiferromagnetic structure. [It is known in the literature that the absence of the upturn in general does not necessarily mean that the magnetic gaps are absent, but the upturn can be overcompensated by the dominance of the competing contribution from the loss of spin disorder part]. Since the second magnetic transition (occurring at 12 K) is very close to $T_N$ for Gd case, this is not resolved in the resistivity data. However, in the case of Tb and Dy, another upturn is clearly observed at the respective second transition temperatures (6.5 K and 2.5 K). Though it is not possible to infer the functional form of $\rho(T)$ as $T \rightarrow 2$ K for Tb and Dy due to the interference from the upturn arising from second transition, one can distinctly see that the functional form appears to be linear for Gd - that is, not quadratic form expected for ferromagnets; the linear behavior - different from $T^5$-dependence predicted for antiferromagnets [36] - further supports complex nature of antiferromagnetic ordering. In the presence of external fields, the values of $\rho$ are reduced in the magnetically ordered state, supporting negative MR seen in the isothermal data. While the upturn at $T_N$ persists for the Tb case even at high fields establishing robustness of magnetic gaps, it shifts towards lower temperature range with increasing $H$, consistent with the conclusion on AFM (at the onset) in zero field from other experiments (see above); the feature due to second magnetic transition is smeared beyond 10 kOe. Similar findings are made for the Dy compound, though the upturn at $T_N$ is smeared at higher fields (say for 70 kOe).

We *now turn to an important finding in the paramagnetic state*. Though the temperature coefficient of $\rho$ is positive, a careful look at the zero-field curve suggests that there is an upward curvature (as demonstrated for the Gd case by drawing a dashed line in Fig. 7) as the temperature is lowered towards $T_N$. This signals that an additional contribution to $\rho$ develops gradually, as the material tends towards magnetic ordering. The temperature where this develops is more than twice of respective $T_N$, unless in systems (e.g., discussed for Dy metal, Ref. 37), where critical point effects can cause such an enhancement of $\rho$ in a smaller temperature window (restricting to a small fraction of magnetic ordering temperature), as argued in Ref. 38. This manifests clearly in the form of a minimum in the case of $Gd_2PdSi_3$, and we had identified a good number of such heavy rare-earth intermetallic compounds in the past [38 - 39]. In other words, the spin-disorder contribution does not appear to be constant in the paramagnetic state in these materials. In order to show that it is truly a spin-related effect, we have measured $\rho$ in the presence of external fields. It is obvious from the figures that the upward curvature is gradually suppressed with an increase of $H$; besides, the curves (in the paramagnetic state) shift downwards with a gradual increase of $H$. These features arise from the fact that the spin-fluctuations (classical spin-liquid phase, as proposed in Ref. 14) get suppressed with magnetic field. The magnitude of MR apparently increases with decreasing $T$ (in other words, increasing MR) as inferred from Fig. 7, as though spin-fluctuations are more pronounced in zero-field as $T_N$ is approached. The magnitude of MR even at twice of $T_N$ is significant (2 to 3%). Needless to add that, usually, in such temperature ranges, MR is negligible for a normal paramagnetic metallic system, with the domination of positive contribution from conduction electrons, as demonstrated for $GdCu_2Ge_2$ (Ref. 38). We had therefore raised a question about two decades ago [38] whether we understand electron correlations for the 'so-called' normal paramagnets (like those of Gd, Tb or Dy in which the Kondo effect is absent). This was a puzzle for a long time. A theoretical advancement emerged [14] only a few years ago and it is interesting that such upturns in spin-disorder contribution as $T_N$ is approached can be explained even within the RKKY interaction incorporating geometrical frustration (leading to a classical spin-liquid phase). We thus believe that the compounds under discussion serve as testing grounds for such novel theories in future.

## IV. SUMMARY

The magnetic and transport behavior of the compounds, $R_2RhSi_3$, containing a geometrically frustrated magnetic lattice, are found to be quite complex. Signatures of a competition between antiferromagnetic and ferromagnetic correlations without spin-glass features, double magnetic transitions, magnetic gap features in ρ, increasing magnitude of (negative) MR with decreasing $T$ towards $T_N$, and large magnetocaloric effect at $T_N$ are major commonalities amongst these Rh compounds. The key point we make is that these properties are quite comparable to those of isostructural $Gd_2PdSi_3$, which is now considered to be a novel magnetic skyrmion. In this connection, a common feature we stress is the observation of two field-induced changes well below $T_N$ in the isothermal data, as in the case of $Gd_2PdSi_3$, as well as in many commonly known field-induced magnetic skyrmions. In view of many resemblances with the properties of $Gd_2PdSi_3$, it is of interest to subject this Rh family for further studies to search for magnetic skyrmions due to geometrically frustrated magnetism. This family may also be a testing ground for the classical spin-liquid phase emerging out of an interplay between RKKY interaction and geometrical frustration, as proposed theoretically by Wang et al [14]. We plan single crystal studies to address these issues.

**Acknowledgment**
One of the authors (EVS) would like to thank Science and Engineering Research Board (Government of India) for supporting this work by awarding J C Bose National Fellowship.

References:

1. See, for a recent review, Bom Soo Kim, J. Phys.: Condens. Matter 31, 383001 (2019).
2. S. Mühlbauer et al., Science **323**, 915 (2009).
3. X.Z. Yu et al., Nature **465**, 901 (2010).
4. E.V. Sampathkumaran, arXiv:1910.09194; also, see the link https://science.sciencemag.org/content/365/6456/914/tab-e-letters
5. R. Mallik E.V. Sampathkumaran, M. Strecker, and G. Wortmann, Europhys Lett. **41**, 315 (1998).
6. R. Mallik et al., Pramana – J. Phys. **51**, 505 (1999).
7. S.R. Saha et al., Phys. Rev. B**60**, 12162 (1999).
8. T. Kurumaji et al., Science **365**, 914 (2019).
9. B. Chevalier, P. Lejay, J. Etourneau, and P. Hagenmuller, Solid State Commun. **49**, 753 (1984).
10. P. A. Kotsanidis, J.K. Yakinthos, and E. Gamari-seale, J. Magn. Magn. Mater., **87**, 199 (1990).
11. R.E. Gladyshevskii, K. Censual, and E. Parthe, J. Alloys and Compd. **189**, 221 (1992).
12. Subham Majumdar et al., J. Phys.: Cond. Matter. **11** (1999) L329.
13. E.V. Sampathkumaran, I. Das, R. Rawat, and Subham Majumdar, App. Phys. Lett. **77**, 418 (2000).
14. Z. Wang, K. Barros, G.-W. Chern, D.L. Maslov, and C.D. Batista, Phys. Rev. Lett. **117**, 206601 (2016); Z.Wang and C.D. Batista, arXiv:2002.03858

15. K. Mukherjee, Tathamay Basu, Kartik K Iyer, and E.V. Sampathkumaran, Phys. Rev. B **84**, 184415 (2011).
16. M. Smidman et al., Phys. Rev. B **100**, 134423 (2019).
17. K. Maiti et al., 1909.10011.
18. I. Das and E.V. Sampathkumaran, J. Magn. Magn. Mater. **137**, L239 (1994).
19. J. Leceijewicz, N. Stüsser, A. Szytula, and A. Szygmunt, J. Magn. Magn. Mater. **147**, 45 (1995).
20. M. Szlawska, D. Kaczorowski, A. Slebarski, L. Gulay, and J. Stepien-Damm, Phys. Rev. B **79**, 134435 (2009).
21. N. Kase, T. Muranaka, and J. Akimitsu, J. Magn. Magn. Mater. 321, 3380 (2009).
22. T. Nakano et al., J. Phys.: Condens. Matter **19**, 326205 (2007).
23. S. Patil, K.K. Iyer, K. Maiti, and E.V. Sampathkumaran, Phys. Rev. B **77**, 094443 (2008); S. Patil et al., Phys. Rev. B **82**, 104428 (2010); S. Patil et al., Eur. Phys. Lett. **97,** 17004 (2012).
24. K. Mukherjee et al., J. Phys.: Conference Series **273,** 012010 (2011).
25. R.-D. Hoffmann and R. Pöttgen, Z. Kristallogr. **216,** 127 (2001).
26. W. Bazela et al., J. Alloys and Comp. **360**, 76 (2003).
27. D.R. Noakes and G.K. Shenoy, Phys. Lett. A **91**, 35 (1982); B.D. Dunlap et al., Phys. Rev. B **29**, 6244 (1984).
28. R. Mallik et al., J. Magn. Magn. Mater. **185**, L135 (1998).
29. D.X. Li et al., Phys. Rev. B **57**, 7434 (1998); D.X. Li et al., J. Phys.: Condens. Matter **11**, 8263 (1999).
30. L.D.C. Jaubert et al., Phys. Rev. Lett. **115**, 267208 (2015).
31. Venkatesh Chandragiri, Kartik K Iyer, and E.V. Sampathkumaran, J. Phys. Condens. Matter **28**, 286002 (2016).
32. Venkatesh Chandragiri, Kartik K Iyer, and E.V. Sampathkumaran, Intermetallics, **76**, 26 (2016).
33. E.V. Sampathkumaran, Kartik K Iyer, Sanjay K Upadhyay, and A.V. Andreev, Solid State Commun. **288**, 64 (2019).
34. Sanjay Kumar Upadhyay, P.L. Paulose, and E.V. Sampathkumaran, Phys. Rev. B **96**, 014418 (2017).
35. M. Bouvier, P. Lethuillier, and D. Schmitt, Phys. Rev. B **43,** 13137 (1991).
36. H. Yamada and S. Takada, Prog. Theor. Phys. **52**, 1077 (1974).
37. R.A. Craven and R.D. Parks, Phys. Rev. Lett. 31, 383 (1973).
38. E.V. Sampathkumaran and R. Mallik, in "Concepts in Electron Correlations", Ed. A.C. Hewson and V. Zlatic, 2003, page 353; R. Mallik and E.V. Sampathkumaran, Phys. Rev. B **58**, 9178 (1998).
39. Ram Kumar, Jyoti Sharma, Kartik K Iyer, and E.V. Sampathkumaran, J. Magn. Magn. Mater **490**, 165515 (2019); Ram Kumar, Kartik K Iyer, P.L. Paulose, and E.V. Sampathkumaran, J. App. Phys. **126**, 123906 (2019).

Figure 1:
Crystal structure of $R_2RhSi_3$ viewed along two different orientations to highlight honeycomb structure and triangular arrangement of R ions.

Figure 2:
Rietveld fitted x-ray diffraction patterns of $R_2RhSi_3$ and difference (bottommost curve for each) between experimental and fitted patterns is also shown. The fitted lines are shown by continuous lines.

Figure 3:
Magnetic susceptibility ($\chi$) measured in a field of 100 Oe (left) and inverse $\chi^{-1}$ measured in 5 kOe (right) for the compounds, $R_2RhSi_3$ (R= Gd, Tb, and Dy). The linear line through the data points represent high-temperature Curie-Weiss region.

Figure 4:
(a) (Top row) Heat-capacity as a function of temperature in zero-field as well as in field, (b) (middle row) 4f-contribution to heat-capacity, and (c) (bottom row) isothermal entropy change, for $R_2RhSi_3$ (R= Gd, Tb, and Dy).

Figure 5:
Isothermal magnetization as a function of magnetic field at selected temperatures for $R_2RhSi_3$ (R= Gd, Tb, and Dy). The derivative curves are shown for 2 K only.

Figure 6:
Isothermal magnetoresistance at 2 K for $R_2RhSi_3$ (R= Gd, Tb, and Dy).

Figure 7:
Zero-field and in-field electrical resistivity as a function of temperature $R_2RhSi_3$ (R= Gd, Tb, and Dy) in the range 2-60 K. In the inset, the zero-field curves extending to higher temperature range are shown to convey that the d$\rho$/d$T$ remains positive in the paramagnetic state at high temperatures. In the mainframe, a line is drawn through the data points well above $T_N$ for the Gd case to show that there is an upward curvature as magnetic ordering is approached; arrows are drawn for the identification of the curves.

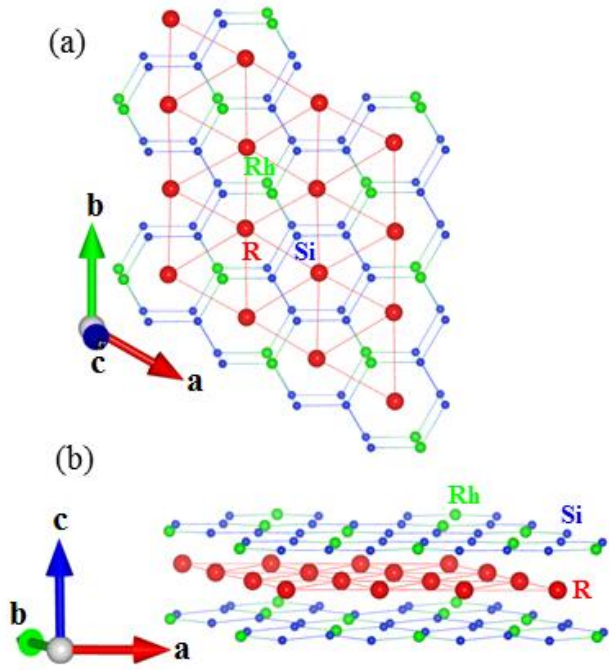

Fig. 1

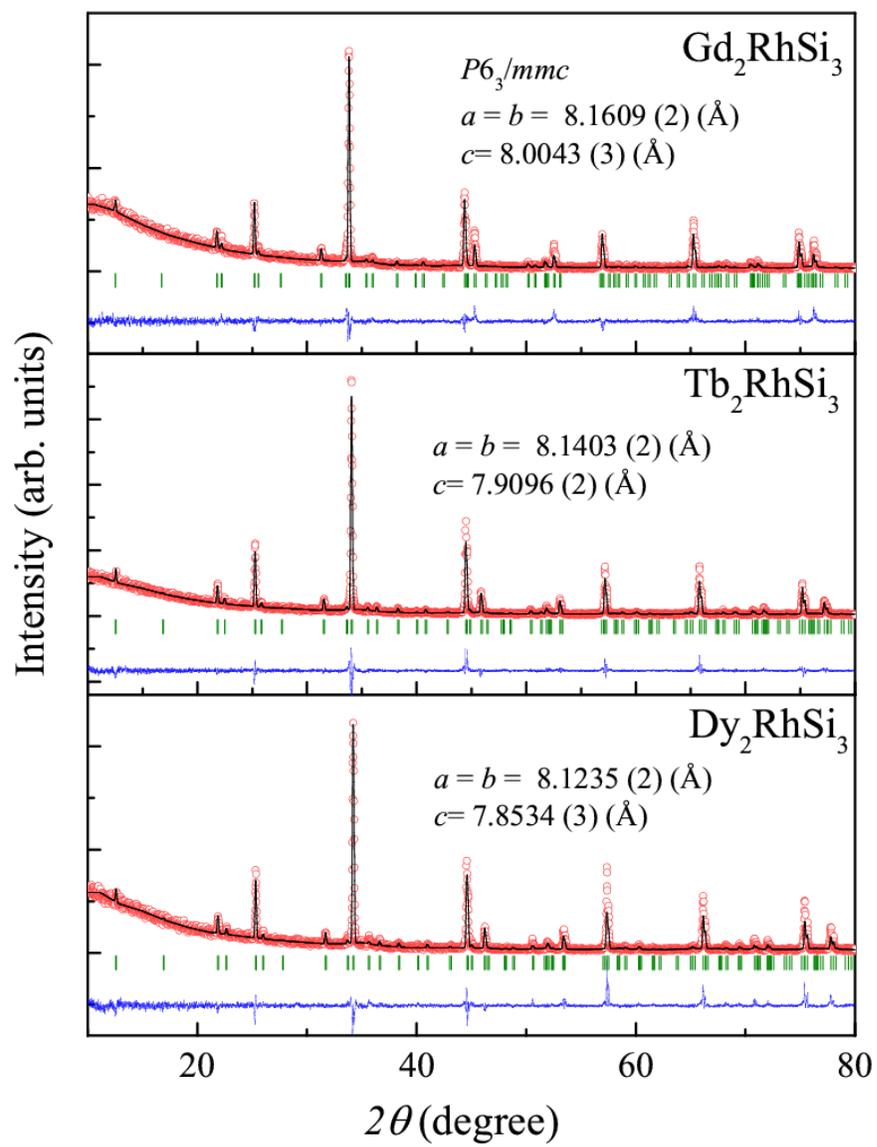

Fig. 2

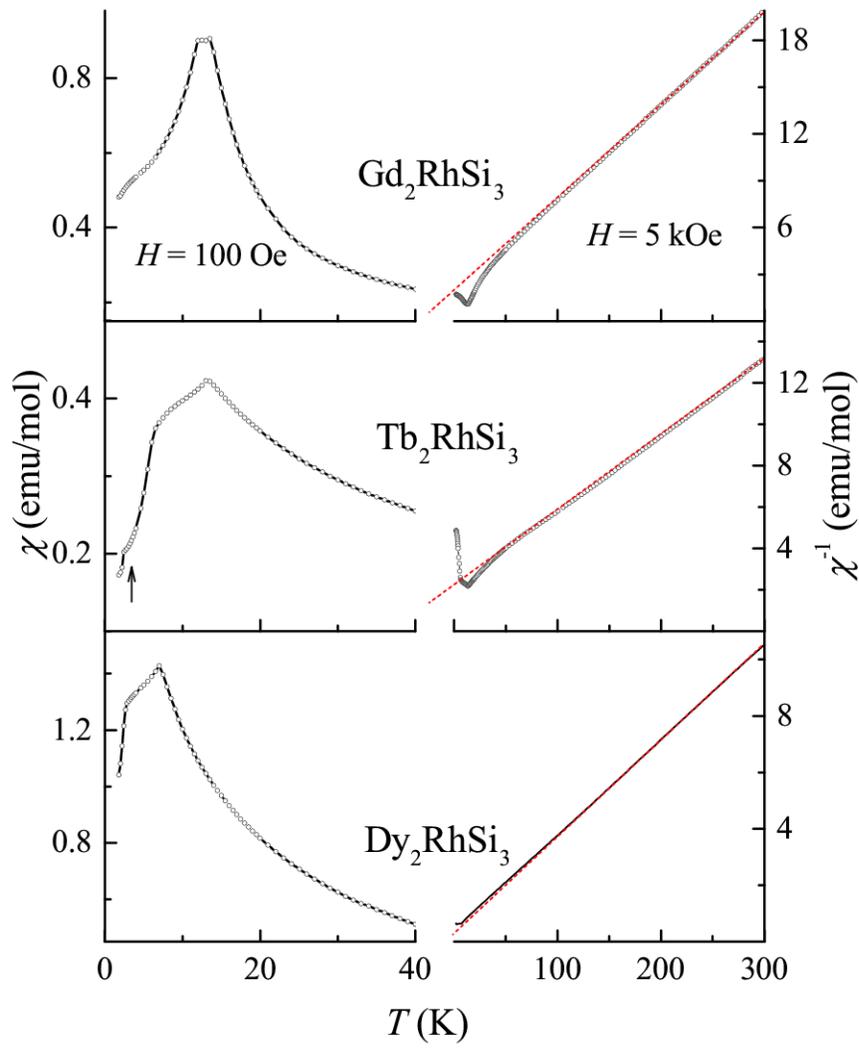

Fig. 3

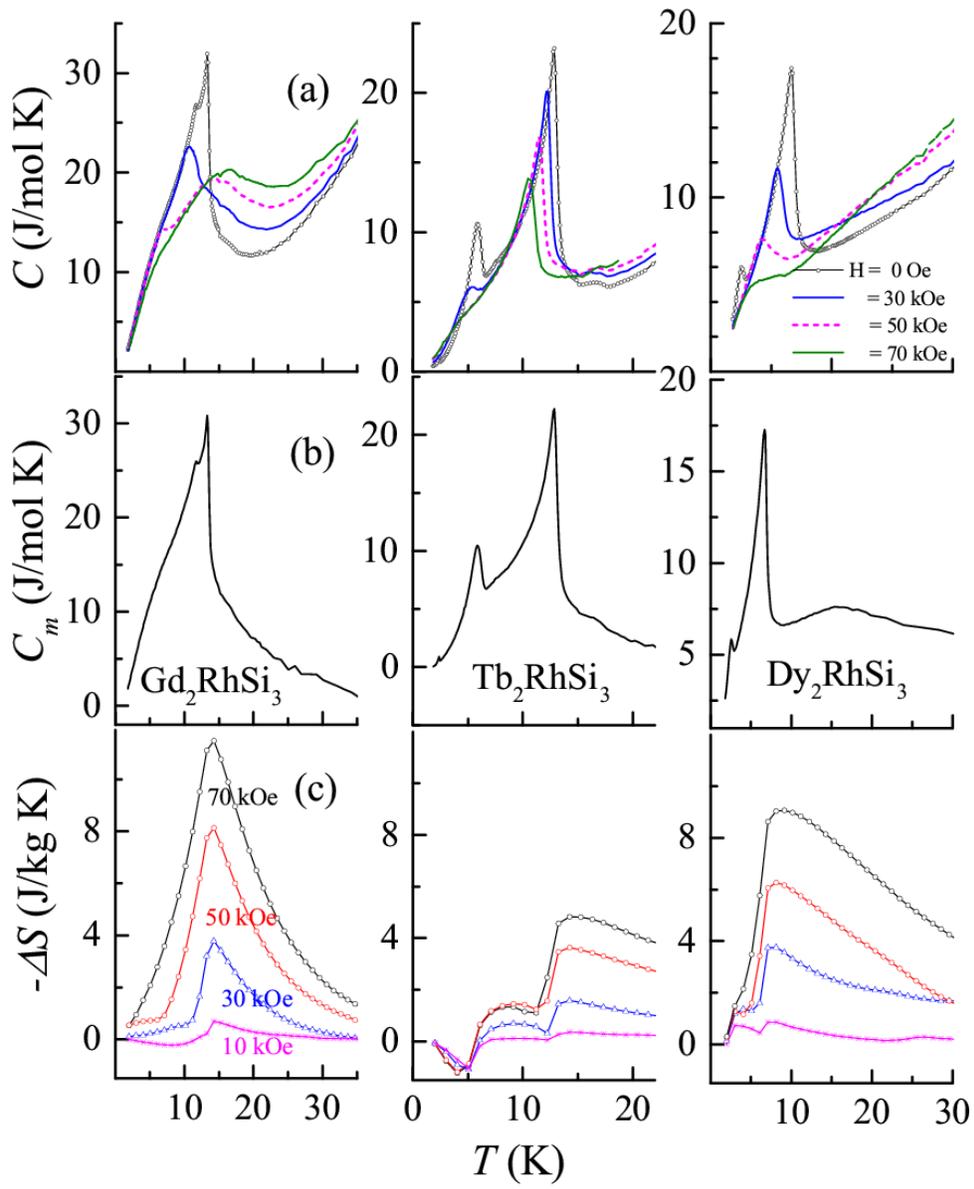

Figure 4

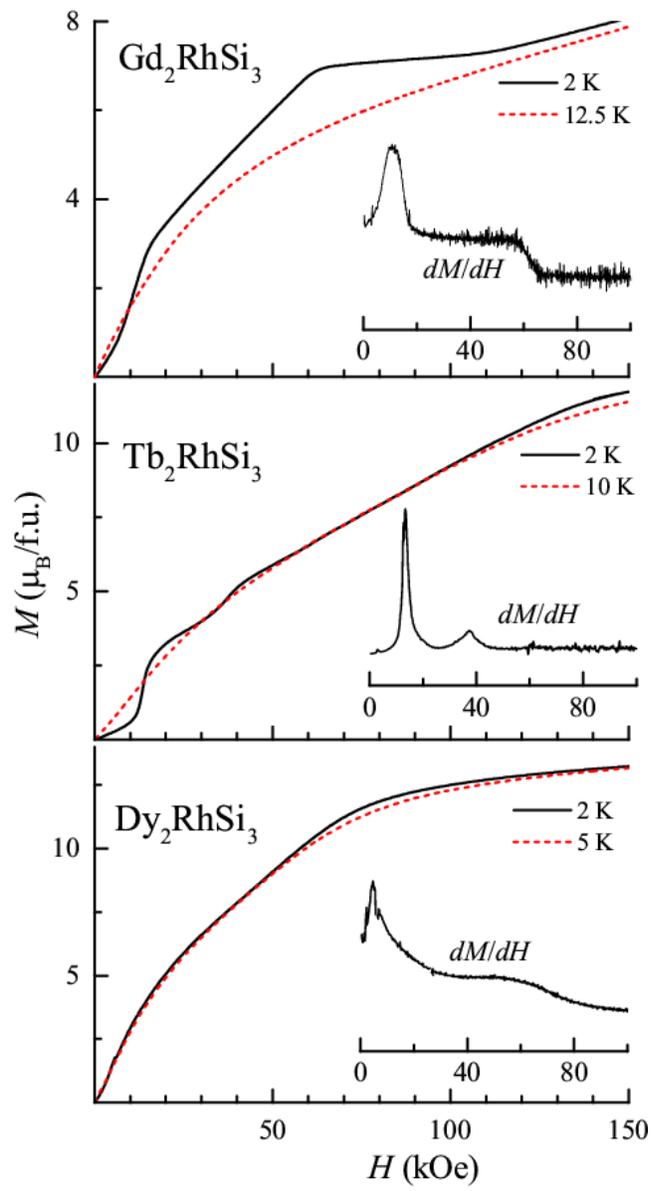

Fig. 5

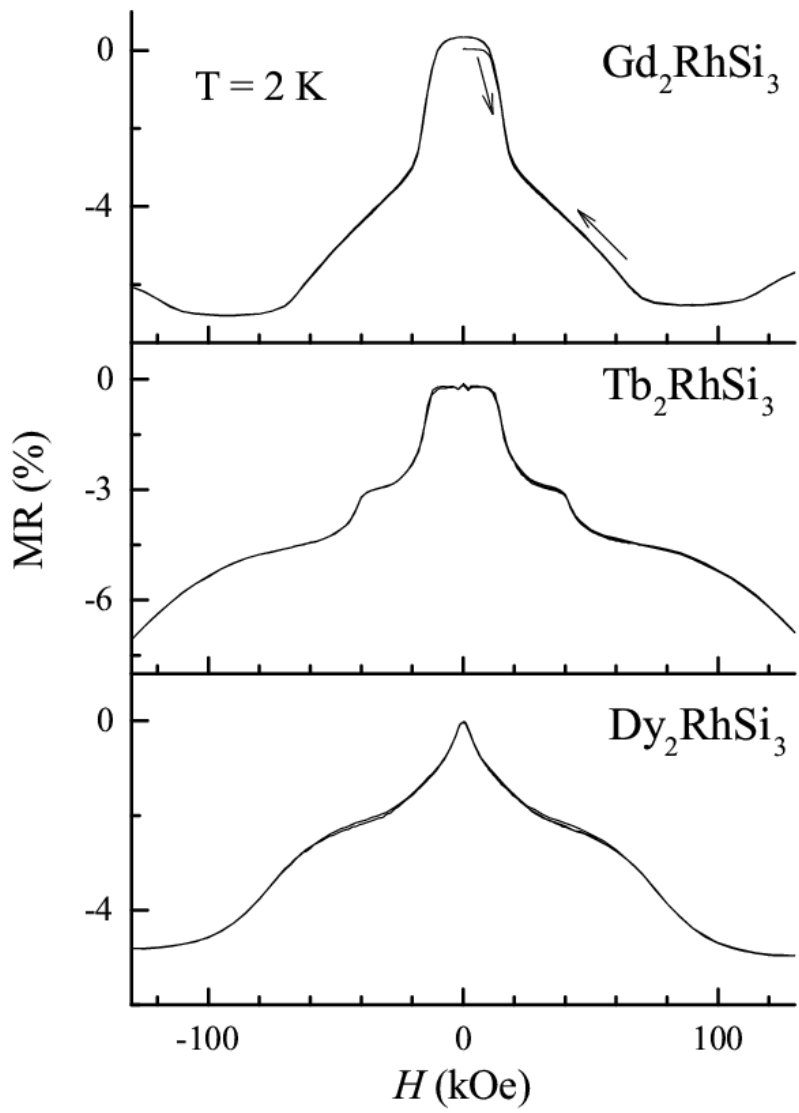

Figure 6

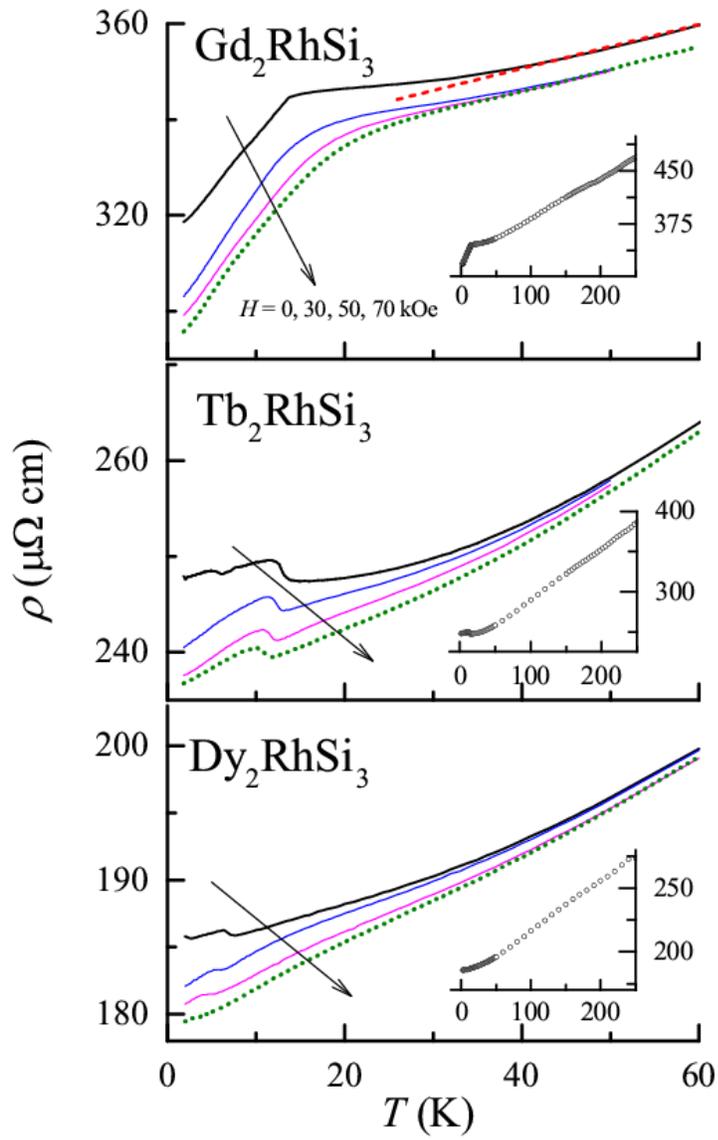

Figure 7